\documentclass[preprint,showpacs,showkeys,preprintnumbers,amsmath,amssymb,pre]{revtex4}

\newcommand{\be}{\begin{equation}}
\newcommand{\ee}{\end{equation}}
\newcommand{\bea}{\begin{eqnarray}}
\newcommand{\eea}{\end{eqnarray}}

\usepackage{latexsym}
\usepackage{color}
\usepackage{graphicx}
\usepackage{dcolumn}
\usepackage{bm}
\usepackage{epstopdf}
\color{black}
\begin{document}
\title{On the Scaling of Langevin and Molecular Dynamics Persistence Times of Non--Homogeneous Fluids.}
\author{Wilmer Olivares--Rivas}
\email{wilmer@ula.ve}
\author{Pedro J. Colmenares}
\email{ colmenar@ula.ve}
\thanks{corresponding author}
\affiliation{Grupo de Qu\'{\i}mica Te\'orica, Quimicof\'{\i}sica de
Fluidos y
 Fen\'omenos Interfaciales (QUIFFIS) \\
 Departamento de Qu\'{\i}mica -- Universidad de Los Andes \\
M\'erida 5101, Venezuela}
\date{\today}
\pacs{02.50.-r; 02.70.-c; 05.10.Gg; 47.10.-g; 61.20.Lc}
\keywords{Molecular Dynamics, Langevin Equation,
 Diffusion constant, Mean Exit Time, mean first passage time, persistence probability}

\begin{abstract}
The existing solution for the Langevin equation of an anisotropic fluid allowed the evaluation of the position dependent perpendicular and parallel diffusion coefficients, using Molecular Dynamics data. However, the time scale of the Langevin Dynamics and Molecular Dynamics are different and an anzat for the persistence probability relaxation time was needed. Here we show how the solution for the average persistence probability obtained from the Smoluchowski-Fokker-Planck equation \textcolor{black} {(SE)}, associated to the Langevin Dynamics, scales with the corresponding Molecular Dynamics quantity. Our  \textcolor{black} {SE} perpendicular persistence time is evaluated in terms of simple integrals over the equilibrium local density. When properly scaled by the perpendicular diffusion coefficient, it gives a good match with that obtained from Molecular Dynamics.
\end{abstract}
\maketitle
\section{Introduction}

The understanding of the dynamics of anisotropic fluids is
fundamental in the study of many chemical and biophysical
processes, particularly those occurring under nano-confinement.
Molecular quantities like the mean square displacement (MSD),
the mean first passage time (MFPT), and probability functions as
 the survival or persistence probability, can be readily obtained
  from a regular Molecular Dynamics (MD) simulation.
  However, {phenomenological quantities are
  bounded to the equations and approximations that defined them.
  Thus, one commonly has to invoke some sort of approximation or condition for their validity}. For instance, the diffusion constant is defined through the stochastic Langevin equation as a ratio of the approximate fluctuation force to the mechanical friction force in the so called Smoluchowski time scale. On a previous work \cite{clor} , we presented an analytical solution for the Langevin equation of an anisotropic fluid, next to an attractive wall,
which allowed the evaluation of the position dependent
perpendicular diffusion coefficient. The mean squared displacement (MSD) and the average persistence time $\tau(z)$, needed to compute this coefficient, were obtained from a modification of the \emph{virtual layers molecular dynamics} method (VLMD) of Liu \emph{et al.}\cite{liu}.\\

A serious difficulty found was that the time scale of the Langevin Dynamics and of the Molecular Dynamics did not match. In fact, it was  discussed by
Burschka \emph{et al}. \cite{burska}, Harris
\cite{harris}, and Razi Naqvi \emph{et al.}
\cite{razi} ,  that the conditional probability obtained as
the stationary solution of the Fokker--Planck equation,
associated to the Langevin equation for a system with
absorbing boundaries conditions fails to vanish at those boundaries. For the same space sampling layers width $L$, the Langevin  persistence probability is
then known to be lower than the corresponding MD persistence
probability. That is, MD particles reach
the boundary faster than \textcolor{black}{Langevin} particles. In another words,
the Langevin dynamics  \textcolor{black}{(LD) }average persistence time, $\tau_{_{LD}}$, for a given sampling layer width, is lower than the corresponding molecular dynamics time
$\tau_{_{MD}}$. {This is a systematic inconsistency, found even for
virtual absorbing layers located at the bulk. It has been handled in the
literature by shifting the boundaries away.} In fact, Liu,
Harder and Berne \cite{liu} overcame this difficulty, by
running simultaneous dual numerical LD and MD simulations, changing the value
of $L$, until the survival probabilities from both
simulations matched. Alternatively we appealed to
the simplest ansatz  of fixing the layer width
and  shifting the $\tau_{_{MD}}$ time instead \cite{clor} .\\

In this paper we take a closer look at this problem, by solving the backwards Smoluchowski-Fokker-Planck equation  \textcolor{black}{(SE)} associated to the anisotropic Langevin equation  and comparing the resulting $\tau_{_{ \textcolor{black}{SE}}}(z)$ with the $\tau_{_{MD}}(z)$ measured from the VLMD method.
In the next section we briefly review the probability concepts in the VLMD method. After establishing the relation between the mean first passage time and the persistence time, we shall obtain an expression for the  \textcolor{black}{SE} average persistence time. Finally in the last section, we shall discuss the scaling between the  \textcolor{black}{SE} and MD dynamics, carrying out calculations for a simple Lennard-Jones dense fluid next to an attractive wall.\\

\section{Virtual Layer Molecular Dynamics Method}

 In this work we will use the VLMD method as discussed previously\cite{clor} . It is a modification of the one introduced by Liu \textit{et al}. (LHB) \cite{liu} and Thomas \textit{et al}. \cite{thomas}.  Briefly, once the dynamics is equilibrated, the total simulation
time $s$ is divided in discrete time steps
$\Delta s$ and partitioned in $J$ blocks containing $n_{max}$ time steps. The MD simulation time interval after n steps is then $t= n\Delta s$. The z coordinate perpendicular to the attractive walls is also divided in discrete virtual layers of width $L$.  We then consider the set of
particles that stay in a given layer located at $z_{a}$, i.e.
$z(t)\in \{a,b\}=\{z_{a},z_{a}+ L \}$, during the
time interval $t$, spanning between the simulation time $s_0$ and $s=s_0 + t$. The initial number of particles in the layer
at $s=s_0$ is $N(0)=N(s_0)$, and the number of particles
in the set, still in the layer after the interval $t$, is
$N(t)=N(s_0 + t)$.   The maximum time interval used to evaluate quantities
inside the layers was $t_{max}=n_{max} \Delta s$. After
$n_{max}$ steps the algorithm is re--initiated to measure
the dynamics in the layer, setting $s_0=s$ again. This
layer sampling is repeated $J$ times. We denote the number
of particles in the set that stay in a layer in the
$j^{th}$ layer
sampling or repetition as $N_j(t)$.\\

 In the VLMD method, the dynamical properties of the anisotropic fluid are evaluated layer by layer using absorbing boundary conditions. That is, particles that exit a given layer are not further counted, even if they reenter the region. The dynamics in the $z$ direction is governed by the behavior of $P(z,t;z_{_{0}})$, the joint probability of finding a particle at position $z_{_{0}}$ at time $t=0$ and then at position $z$ at time $t$. For an anisotropic system this function depends on the position where the layer is located, a$=z_{_{a}}$, and on the width $L$ of the space interval. \textcolor{black}{The validity and accuracy of the VLMD
 method rely on the proper choice of  L\cite{clor,liu} . It was found that if L is too large the mean force within the layer changes more than required by the method. While, if it is too small, particles tend to escape from the virtual layer, affecting the proper statistics of the MD. For the fluid conditions used, L equal to 0.5 diameters was found to be a good compromise.}\\
 
 Since the movement in the $z$  direction is assumed to be independent of the movement in the $x$ and $y$ directions, the joint probability function is given as
 
\be P(z,t,z_{_{0}}) = P(z,t|z_{_{0}}) p_{a}(z_{_{0}}),\label{eqjoinP}\ee
where $P(z,t|z_{_{0}})$ is the  bayesian conditional probability
that the particle is located at $z$ at time $t$, given
that it was at $z_{_{0}}$ at time $t=0$, in a layer
containing $z_{_{0}}$. $p_{a}(z_{_{0}})$ is the probability
for a particle to be at position $z_{_{0}}$ at time $t=0$, normalized in the given region. It can be obtained directly from the local particle density $\rho(z)$.
\be
 p_{a}(z_{_{0}})=\frac{\rho(z_{_{0}})}{\int_{a}^{b}\rho(z)dz}.
 \label{eqprobz0}
\ee

The normalization of the joint probability is by definition the \emph{persistence
probability}, $\mathcal{P}(t,z_{a})$, in the diffusion domain $z_{a} \le z < z_{a}+L$
 \be
 \mathcal{P}(t,z_{a}) = \int_{z_{a}}^{z_{a}+L}\!\!\!dz\int_{z_{a}}^{z_{a}+L}\!\!\!dz_{_{0}}
 P(z,t;z_{_{0}}).
 \label{eqP(t)}
 \ee

 It measures the average
probability that after a time interval $t$ a particle
still remains inside a layer located at $z_{a}$,
 independently of the initial position.
 $\mathcal{P}(t;z_{a})$ is often also called the
 survival probability \cite{liu} , however, we prefer to call it
 the persistence  probability to distinguish from the \emph{survival probability}
  $G(t,z_{_{0}};z_{_{a}})$, as defined
  {in the literature}
  \cite{gardiner,ColmenaresPRE}
  \be
G(t,z_{_{0}};z_{_{a}})=\int_{a}^{b}P(z,t|z_{_{0}})dz. \label{eqG}
\ee

$G(t,z_{_{0}};z_{_{a}})$ is the probability that at time $t$, the particle, initially
located at $z_{_{0}}$, remains anywhere in the region
$\{a,b\}$. So, while the survival probability is a function of the initial position, the persistence probability  is independent of it. \\

In a MD simulation, the probability $\mathcal{P}(t,z_{a})$ is obtained by averaging
$P_{j}(t)=N_j(t)/N_j(0)$ over the $J$ repetitions
\be
\mathcal{P}(t,z_{a})=\frac{1}{J} \sum_{j=1}^J P_{j}(t).
\label{eqP(t)MD}
\ee

As it has been shown by Molecular, Generalized Langevin and Langevin Dynamics $\mathcal{P}(t,z_{_{a}})$ can be phenomenologically approximated with a high degree of accuracy by an exponential
decay \cite{liu,thomas,clor} ,
\be
\mathcal{P}(t,z_{_{a}})=e^{^{-t/\tau(z_{_{a}})}},
\label{P(t)fromEXP}\ee
 where $\tau(z_{_{a}})$ is the the relaxation time of the persistence process, also referred to as the \emph{average persistence time}. The MD $\tau(z_{_{a}})$ is evaluated from Eq. (\ref{P(t)fromEXP}), by  fitting the  numerical $\mathcal{P}(t,z_{_{a}})$ MD data.\\

All other physical properties of interest are evaluated as averages over the joint probability. For instance, the mean square displacement (MSD) in a layer is obtained as the average
\bea
 \left\langle(z(t)-z_{_{0}})^2\right\rangle_{a}= \int_{z_{a}}^{z_{a}+L}\!\!\!dz\int_{z_{a}}^{z_{a}+L}\!\!\!dz_{_{0}} [z(t)-z_{_{0}}]^{^{2}}P(z,t,z_{_{0}}).
 \label{eq34}
\eea
This expression also holds
for the $x$ and $y$ directions. In the MD simulation the mean square
displacement (MSD) within the layer at $a$ is
evaluated summing over all the $N_{j}(t)$ particles in the
set

\be
\left\langle\left[z(t)-z(0)\right]^{2}\right
\rangle_{a}
=  \frac{1}{J} \sum_{j}^{J}\frac{1}{P_{j}(t)}\left[\frac{1}{N_j(0)}
\sum_{i}^{N_j(t)}\left[z_i(t)-z_i(0)\right]^{2}\right].
\label{avMSD}
\ee

\section{Mean First Passage Time  and Persistence Probability}
In this section we will develop the relation between the
MFPT and the mean survival time of particles dwelling in
the layer. Using Eqs. (\ref{eqjoinP}) and (\ref{eqG}),
we rewrite Eq. (\ref{eqP(t)}) as an average over the initial position and find that a particle dwells in a given layer with a persistence
probability
\bea
  \mathcal{P}(t;z_{a})&=& \int_{z_{a}}^{z_{a}+L}\!\!\!dz\left[\int_{z_{a}}^{z_{a}+L}\!\!\!dz_{_{0}}
 P(z,t|z_{_{0}})\right]p_{_{a}}(z_{_{0}}),
 \label{eqP(t)2}\\
  &=&
\int_{a}^{b}dz_{_{0}}G(z_{_{0} },t) p_{_{a}}(z_{_{0} }) .\label{P(t)fromG}\eea

 In the context of this note, the average persistence time that the  particles spend in motion within the virtual layer is then given, according to Eq. (\ref{P(t)fromEXP}) by the integration  of $\mathcal{P}(t,z_{_{a}})$ over the time interval. Using Eqs. (\ref{eqP(t)2}) and (\ref{P(t)fromG}) we get \bea
\tau(z_{_{a}})&=&\int_{0}^{\infty}dt \mathcal{P}(t,z_{_{a}}),\nonumber \\
&=&\int_{a}^{b}dz_{_{0}}\left[\int_{0}^{\infty}dtG(z_{_{0} },t) \right] p_{_{a}}(z_{_{0} }),\nonumber\\
&=&\int_{a}^{b}dz_{_{0}} [t_{_{MFP}}(z_{_{0}})]p_{_{a}}(z_{_{0} }),\label{tauFROMtMFP}
\eea
where we have defined the mean time it takes a
particle to reach the boundary given that it
started at $z_{_{0} }$ as the
\emph{mean first passage time}
$t_{_{MFP}}(z_{_{0}})$ (MFPT){\cite{gardiner}}

\be
t_{_{MFP}}(z_{_{0}})=\int_{0}^{\infty}dtG(z_{_{0} },t).
\label{eqMFPTfromG}\ee

Note that when the virtual boundaries of the layer are
{absorbing,} the MFPT is actually a \emph{mean exit time}.
Thus, according to Eq. (\ref{tauFROMtMFP}), $\tau(z_{_{a}})$ is simply the space average of $t_{_{MFP}}(z_{_{0}})$
\be
\tau(z_{_{a}}) =\langle t_{_{MFP}}(z_{_{a}})\rangle_{_{a}}.
\label{tau(a)}\ee

The MFPT have the nice feature that they can be evaluated in terms of the equilibrium particle density. In earlier work, we have associated the MFPT with effective diffusion constants for confined ionic and non ionic fluids, and with local diffusion coefficients in anisotropic nano--confined molecular dense fluids\cite{clor,ColmenaresPRE,SOVCLTheochem,SulbaranCMP} . So, in the next section we will obtain the expressions for the MFPT from the Smoluchowki-Fokker-Planck equation associated to the Langevin equation.

\section{Langevin Dynamics Mean First Passage Time}
The $z$ component of the anisotropic Langevin--like
equation, can be written as\cite{liu,clor}
\be
m\frac{d\mbox{v}(z)}{dt}=-\gamma_{_{zz}}\mbox{v}(z) + F(z)+
\sigma(z)\,\xi(t)
\label{Langevin}.
\ee
It describes the dynamics of a fluid particle
of mass $m$ located at position $z(t)$ with velocity
$\mbox{v}(z)$, in the presence of and external force
$F(z)$ and a random or fluctuation force $\sigma(z) \xi(t)$. $\gamma_{_{zz}}$ is the z component of the
 friction tensor of the fluid and,
$\xi(t)$ is the usual
$\delta$--correlated--zero--mean white noise, resulting
from the collisions with the rest of the fluid. The coefficient $\sigma(z)$ measures the amplitude of the fluctuation force. In terms of a potential of
 mean force $W(z)$, felt
by the particles due to the interactions with the walls of
the container or and external field, the force is given by
$F(z)=-\frac{dW(z)}{dz}=\textcolor{black}{k_{_{B}}T}\frac{d\ln \rho(z)}{dz}$, where $\rho(z)$ is the local particle density. \textcolor{black}{Here, $k_{_{B}}$ and $T$ are the
Boltzmann constant and temperature,
respectively}. {\color{black}To write Eq. (\ref{Langevin}), we used the standard assumption that the friction tensor, $\bf\Upsilon$, is diagonal with $\gamma_{zz}(z)\neq\gamma_{xx}(z)=\gamma_{yy}(z)$ \cite{liu,clor}}, neglecting the off--diagonal terms $\gamma_{xz}(z) =\gamma_{yz}(z)=0$. Here we used $\gamma(z)$  for {\color{black} the perpendicular or  transverse, $zz$ diagonal element of the friction matrix, $\gamma(z)=\gamma_{_{zz}}$. It has been shown that the parallel $xx$ and $yy$ components are also dependent on the z position, but to a much lesser extent\cite{clor}. Therefore, we shall only consider in this paper the transverse component of the anisotropic diffusion coefficient.}

 In what follows, we suppose the velocities of the particles
attain a canonical distribution
much faster than positions. Thus, at the time scale of
positions, velocities attain the steady state, $\frac{m}{\gamma_{_{zz}}} d\mbox{v}(z)/dt=0$, so the instantaneous relaxation approximation (IRA) \cite{ake} can be applied to
Eq. (\ref{Langevin}) to get
\be
\frac{dz(t)}{dt}= \mbox{v}_{_{\gamma}}(z)+\sqrt{2D(z)}\,\xi(t),
\label{LEira}
\ee
where we defined the space dependent diffusion coefficient $D(z)=D_{_{zz}}$, to relate the fluctuation and the dissipation forces, as $D(z)=\frac{1}{2}(\frac{\sigma(z)}{\gamma(z)})^{2}$ and, we introduced the drift velocity $\mbox{v}_{_{\gamma}}(z)= \frac{F(z)}{\gamma(z)}$. {\color{black} The validity of this approximation, even for ionic fluids, was successfully tested by J\"{o}nsson and Wennestr\"{o}m\cite{jonsson}}. Equation (\ref{LEira}) gives the velocity of the
particle under a stochastic potential field.  This is also referred to as the high friction limit approximation and extensibly used in the literature \cite{gardiner,risken}. Thus, if the
stochastic differential equation (\ref{LEira}) is
interpreted in Ito's sense \cite{gardiner} , the
corresponding forward Fokker--Planck (FP)
equation for the conditional probability
$P(z,t|z_{_{0}})$=$P(z,t|z_{_{0}},t_{_{0}}=0)$ of finding
the particle in position $z=z(t)$ at time $t$, given that
it started to diffuse from an initial position $z_{_{0}}$
at $t_{_{0}}=0$ is\cite{risken}
\be \frac{\partial
P(z,t|z_{_{0}})}{\partial t}=\frac{\partial^{\,2}}{\partial z^{2}}\left[
D(z)P(z,t|z_{_{0}})\right] -\frac{\partial} {\partial z}
\left[\mbox{v}_{_{\gamma}}(z)P(z,t|z_{_{0}}) \right].
\label{fpe} \ee

When one postulates that, at equilibrium, the \textcolor{black}{local} particle density, \textcolor{black}{
written as a Boltzmann distribution $\rho(z)=\rho_{_{B}}\exp(-\beta W(z))$}, is  to be a stationary solution of \textcolor{black}{the FP equation, Eq. (\ref{fpe}),} and that there is no net flux of particles, one gets the \emph{fluctuation-dissipation theorem} in the form
\bea
\frac{d D(z)}{dz} &=& \beta F(z)\left[ \frac{k_{_{B}}T}{\gamma(z)} - D(z) \right],\\
 &=& \mbox{v}_{_{\gamma}}(z)-\mbox{v}_{_{D}}(z),
\label{eqFDT}
\eea
where $\mbox{v}_{_{D}}(z)= \beta F(z) D(z)$, and as above, $\mbox{v}_{_{\gamma}}(z)=\beta F(z)\left[ \frac{k_{_{B}}T}{\gamma(z)}\right]$ with $\textcolor{black}{\beta=1/k_{_{B}} T}$. Since we are interested in anisotropic systems, where $\frac{d D(z)}{dz}$ is nonzero, in general $\mbox{v}_{_{\gamma}}(z) \neq \mbox{v}_{_{D}}(z)$. Therefore, the Sutherland-Einstein relationship, $D(z) = \frac{k_{_{B}}T}{\gamma(z)} $, does not apply. Instead, substituting Eq. (\ref{eqFDT}) in the FP equation, gives the so called  Smoluchowski-Fokker-Planck (SFP) equation in terms of $D(z)$ and $\mbox{v}_{_{D}}(z)$,  which can be written as a continuity equation in the form

\be
\frac{\partial
P(z,t|z_{_{0}})}{\partial t}=-\frac{\partial  j(z,t|z_{_{0}})}{\partial z},
\label{eqSmoluchowski}
\ee
with the flux $j(z,t|z_{_{0}})$  given as
\be
j(z,t|z_{_{0}})=  \mbox{v}_{_{D}}(z)P(z,t|z_{_{0}})-D(z)\frac{\partial P(z,t|z_{_{0}})}{\partial z}.
\label{eqflux}
\ee
\textcolor{black}{This is also commonly referred to as the forward Smoluchowski equation (SE).} \textcolor{black}{Particle dynamics and fluctuations occurring in biological and liquid environments are often described well by this equation\cite{szabo} . The main objective of this paper is to establish the time scale associated to this equation, when compared to MD.}\\

Defining the Smoluchowski operator
\be
\mathcal{L}(z) = \frac{d}{dz}\left[ \rho(z) D(z)\frac{d}{dz}\frac{1}{\rho(z)}\right],
\ee
and its adjoint
\be
\mathcal{L}^{*}(z_{_{0}}) = \frac{1}{\rho(z_{_{0}})} \frac{d}{dz_{_{0}}} \rho(z_{_{0}})D(z_{_{0}})\frac{d}{dz_{_{0}}},
\ee
one can rewrite the previous forward \textcolor{black}{SE} as
\be \frac{\partial
P(z,t|z_{_{0}})}{\partial t}=\mathcal{L}(z)P(z,t|z_{_{0}}),
\label{fSE}
\ee
and the corresponding adjoint or backward  \textcolor{black}{SE}
\be \frac{\partial
P(z,t|z_{_{0}})}{\partial t}=\mathcal{L}^{*}(z_{_{0}})P(z,t|z_{_{0}}),
\label{b-SE}
\ee
where the operator $\mathcal{L}^{*}(z_{_{0}})$ acts on the initial position $z_{_{0}}$.
In the problem
of first passage times, fully  described elsewhere
\cite{gardiner,ColmenaresPRE} , the objective is to know how long a
particle, whose position is described by the above equations, remains in the
region $\{a,b\}$.

From the definition given in Eq. (\ref{eqG}), a simple integration of Eq. (\ref{b-SE}) over $z$, gives a differential equation for the  the survival probability
\be
\frac{\partial
G(z_{_{0}},t)}{\partial t}=\mathcal{L}^{*}(z_{_{0}})G(z_{_{0}},t),
\label{eqG-S}
 \ee
 which must to be
solved with the initial condition $G(z_{_{0}},t=0)=1$.  Carrying out the same integration over both sides of the continuity equation, Eq. (\ref{eqSmoluchowski}), we get
\be
\frac{\partial
G(z_{_{0}},t)}{\partial t}=j(a,t|z_{_{0}}) - j(b,t|z_{_{0}}).
\label{eqGflux}
 \ee

 The right hand side gives the flow out of the layer.
From this, it can be seen that the first passage process to the absorbing layers is in fact an exit process.

The mean exit time or mean first passage time
 is  then evaluated, from Eq. (\ref{eqMFPTfromG}), by integrating the evolution equation for
$G(z_{_{0}},t)$, Eq. (\ref{eqG-S}). So, the MFPT satisfies the differential equation
\be
\mathcal{L}^{*}(z_{_{0}})t_{_{MFP}}(z_{_{0}})=-1.
\label{eqtMFP}
 \ee

The boundary conditions for these equations must fulfill the requirement that the virtual layers are absorbing at $a$ and $b$, in the sense that particles are counted out once they reach the boundary. While some difficulty might exist when dealing with the probability function $P(z,t|z_{_{0}})$  in the  \textcolor{black}{SE} \cite{liu} and even in the stochastic free diffusion equation \cite{mcquarrie, razi,harris} , the absorbing boundary condition for a survival or exit probability is straightforward, namely, $G(a,t)=G(b,t)=0$, and for the MFPT, $t_{_{MFP}}(a)=t_{_{MFP}}(b) =0$.\\

The solution of the differential equation for the MFPT with absorbing boundary conditions follows
\be
t_{_{MFP}}(z_{_{0}})=\frac{\left(\int_{a}^{z_{_{0}}}\frac{dy}{\psi(y)}\right)
\int_{z_{_{0}}}^{b}\frac{dz}{\psi(z)}\int_{a}^{z}\,
\rho(x)dx
-\left(\int_{z_{_{0}}}^{b}\frac{dy}{\psi(y)}\right)
\int_{a}^{z_{_{0}}}\frac{dz}{\psi(z)}\int_{a}^{z}\,
\rho(x)dx}
{\int_{a}^{b}\frac{dz}{\psi(z)}}.
\label{eqMFPT}
\ee
Here, the potential of mean force was related to the local particle density $\rho(z_{_{0}})$ and we defined the extensive diffusion density function $\psi(z_{_{0}})=\rho(z_{_{0}})D(z_{_{0}})$.\\

  Following the VLMD approximations, we now assume that the layer
is small enough that the property $D(z)$ is constant
for $z\in\{a,b\}$. That is, we assume that the Sutherland-Einstein relationship holds locally in the virtual layer, so that $\frac{kT}{\gamma(z)}=D(z)=D(z_{_{a}})$, where the constant $D(z_{_{a}})$ is the diffusion coefficient evaluated at the left boundary of the virtual layer at position $z_{_{a}}$. \\

From Eq. (\ref{tau(a)}), we identify the relaxation
parameter $\tau_{_{ \textcolor{black}{SE}}}$ of the  persistence probability,
as the position--averaged MFPT.
 For an anisotropic system this  position average depends on the position $a=z_{_{a}}$ where the region is located and, on the width $L$ of this space interval.\\
 We finally  get
\be
\tau_{_{ \textcolor{black}{SE}}}(z_{_{a}})=\frac{\int_{a}^{b}\,dz_{_{0}} \rho(z_{_{0}})\left[\left(\int_{a}^{z_{_{0}}}\frac{dy}{\rho(y)}\right)
H[z_{_{0}},b]-\left(\int_{z_{_{0}}}^{b}\frac{dy}{\rho(y)}\right)
H[a,z_{_{0}}]\right]}{D(z_{_{a}})\,\left(\int_{a}^{b}\rho(z)dz\right)\left(\int_{a}^{b}\frac{dz}{\rho(z)} \right)},
\label{eqavMFPT}
\ee
where the integral $H[y,s]$ is
\be H[y,s]=
\int_{y}^{s}\frac{dz}{\rho(z)}\int_{a}^{z}\,
\rho(x)dx.
\ee

Analytical results can be drawn from the previous equations whenever the functional form of the local
particle density $\rho(z)$ is known. A limiting special case is obtained for a region located on the bulk reservoir away from the walls, where the potential of mean force
vanishes. Then, using $\rho(z_{_{0}})=\rho_{_{bulk}}$ in Eq.
(\ref{eqMFPT}) and $p_{a}(z_{_{0}})=1/L$ in Eqs. (\ref{tauFROMtMFP}) and (\ref{eqavMFPT}) we get
\bea
t^{bulk}_{_{MFP}}(z_{_{0}})&=&\frac{(z_{_{0}}-a)(b-z_{_{0}})
}{2D^{bulk}},\\
\tau^{bulk}_{_{ \textcolor{black}{SE}}}&=&\frac{L^{2}}{12D^{bulk}}, \label{eqtau1} \eea where
$D^{bulk}=D\!_{_{0}}$ is the homogeneous
bulk fluid diffusion constant.
It is easily evaluated, for example,
from the long time limit of the MSD in a regular MD simulation.
\section{Results and Discussion}

We studied the same anisotropic system as in reference [1], namely
a dense Lennard--Jones fluid
with a bulk reduced density of $\rho_{bulk}=0.69$, using $1296 $ molecules, at a reduced temperature of $T^{*}= k_{B}T/\epsilon_{FF}=0.75$, next to
a highly
  interacting
  9--3 LJ smeared wall acting with a fluid-wall potential depth of $\epsilon_{FW}=1.0 k_{B}T$
  \cite{israelachvili,Travis&Gubbins} . The bulk was consider in $z$ regions away from the walls, at
$z_{a}=20\sigma$. So Eq. (\ref{eqtau1}) allows the evaluation $\tau^{bulk}_{_{ \textcolor{black}{SE}}}$ for a layer of width $L$.
 The corresponding molecular dynamics
 $\tau^{bulk}_{_{MD}}$ is evaluated {by}
 fitting the persistence probability {curve} with
 Eq. (\ref{P(t)fromEXP}). \textcolor{black}{This fitting was good for all distances $z$ and even for different values of $L$\cite{clor}}.
 
  For a typical simple Argon like fluid with a
 $D^{bulk}$ of $4.0676 \,\,\mbox{x}\,\, 10^{-9}\,\,m^{2}/s$,
 for a layer of width $L=0.5\sigma=0.1668\,\,nm$,
 we get a value $0.57002\,\,ps$ for
 $\tau^{bulk}_{_{ \textcolor{black}{SE}}}$, while the corresponding
 $\tau^{bulk}_{_{MD}}$ from the MD simulation
 is $1.3678\,\,ps$.
Therefore, in the bulk the MD persistence time
is more than twice larger that the  \textcolor{black}{SE} time.
From this results, Eqs. (\ref{eqavMFPT}) and (\ref{eqtau1}) provide a procedure to infer about $\tau_{_{ \textcolor{black}{SE}}}$ simply by knowing the local equilibrium particle density inside the layer. Rewriting Eq. (\ref{eqtau1})
\be D^{bulk}\,\tau^{bulk}_{_{ \textcolor{black}{SE}}} = D^{*}\,\tau^{bulk}_{_{MD}}
 = \frac{L^{2}}{12},\ee
where the
constant $D^{*}=\frac{L^{2}}{12 \tau^{bulk}_{_{MD}}}$.

\begin{figure}[ht]
\includegraphics[height=11cm,width=15.5cm,angle=0]{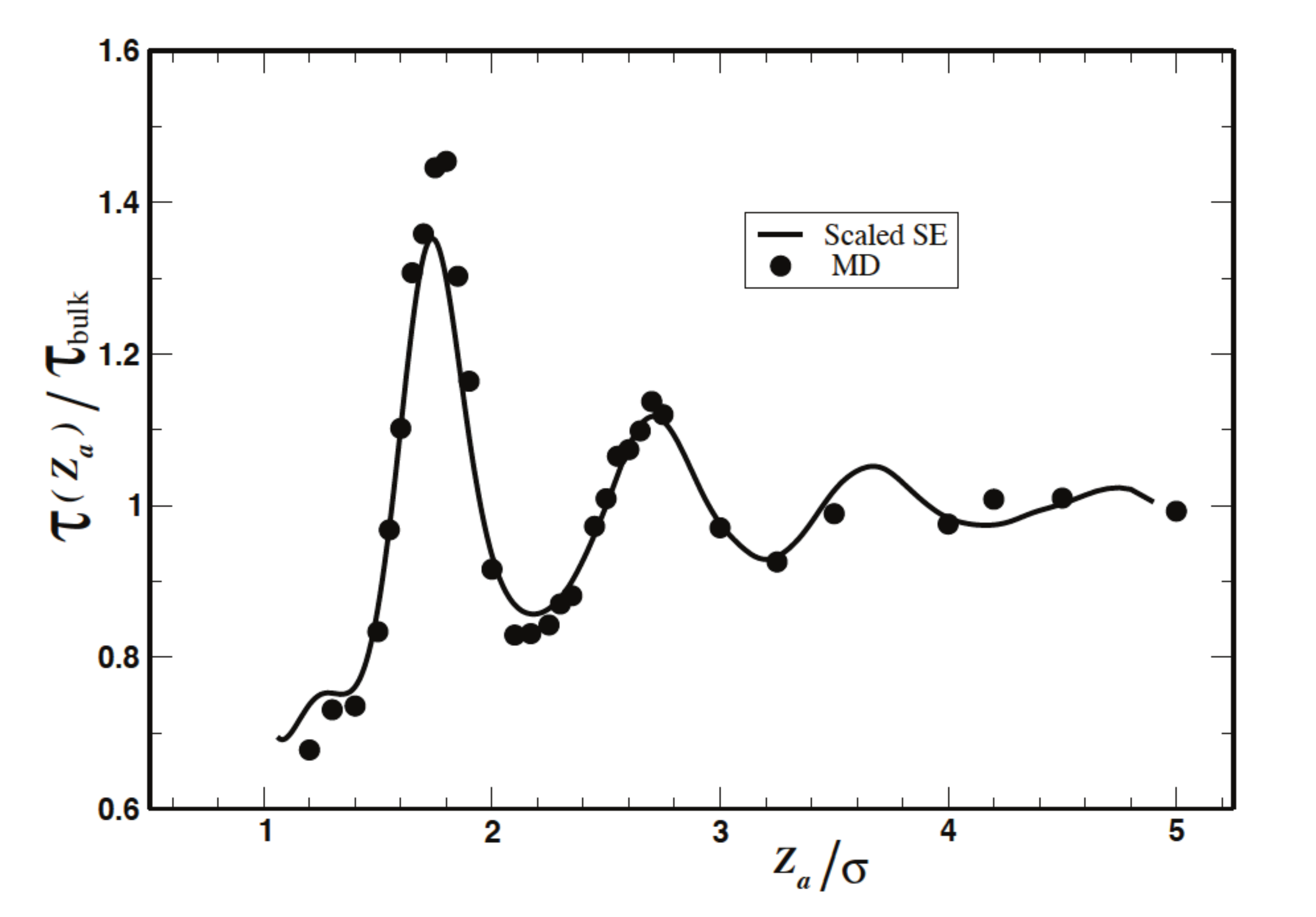}
\caption{\label{Fig1} Average persistence time relative to the corresponding bulk value, as a function of layer position $z_{a}$. $L=0.5\sigma$. Dots are the result from the MD simulation $\left(\frac{\tau_{_{MD}}(z_{a})}{\tau^{bulk}_{_{MD}}}\right)$
\cite{clor} . The continuous line is the result of scaling $\left(\frac{\tau_{_{ \textcolor{black}{SE}}}(z_{a})}{\tau^{bulk}_{_{ \textcolor{black}{SE}}}}\right)$ by the factor $\left[\frac{D(z_{a})}{D^{bulk}}\right] $ according to Eqs. (\ref{tau(a)}) and (\ref{eqavMFPT}).}
\end{figure}

As discussed in \cite{clor} , the numerical MD value of
$\tau^{bulk}_{_{MD}}$ gives a value of $D^{*}=1.69512\,\,\mbox{x}\,\,10^{-9}\,\, m^{2}/s
$ which
differs from the expected value of the bulk diffusion
constant $D^{bulk}$. This is due to the known difference
in the MD and  \textcolor{black}{SE} average persistence probabilities and, by
definition, of $\tau_{_{ \textcolor{black}{SE}}}$ and $\tau_{_{MD}}$, for the
same value of the virtual layer width $L$. Since this is a systematic difference, we  \textcolor{black}{conjecture} that this relation holds also for layers located at
$z_{a}$ in the vicinity of an attractive wall
\bea
D(z_{a})\tau_{_{ \textcolor{black}{SE}}}(z_{a})&=&
 D^{*} \tau_{_{MD}}(z_{a}),\nonumber\\
&=&
\left[
\frac{L^{2}}
{12}\right]
\left(\frac{\tau_{_{MD}}(z_{a})}
{\tau^{bulk}_{_{MD}}}\right),\nonumber\\
&=& \left[ D^{bulk}\,\tau^{bulk}_{_{ \textcolor{black}{SE}}} \right]\left(
\frac{\tau_{_{MD}}(z_{a})}{\tau^{bulk}_{_{MD}}}\right).
\eea
This can be rearranged as
\be
\left(\frac{\tau_{_{MD}}(z_{a})}{\tau^{bulk}_{_{MD}}}\right)
=\left(\frac{\tau_{_{ \textcolor{black}{SE}}}(z_{a})}
{\tau^{bulk}_{_{ \textcolor{black}{SE}}}}\right) \left[\frac{D(z_{a})}{D^{bulk}}\right]. \label{eqScaling}
\ee

This relationship is our main result. The left hand side of this relationship corresponds to a quantity obtained directly from the persistence probability in the Molecular Dynamics simulation, using Eqs. (\ref{eqP(t)MD}) and (\ref{P(t)fromEXP}). The right hand side of Eq. (\ref{eqScaling}) is a quantity that can be evaluated independently from the solution of the  \textcolor{black}{Smoluchowski} equation, Eq. (\ref{eqavMFPT}). It contains the phenomenological diffusion coefficient, introduced in the Langevin equation, Eq. (\ref{Langevin}), in the high friction limit, Eq. (\ref{LEira}), to relate the fluctuation and dissipation forces,  $D(z)=\frac{1}{2}(\frac{\sigma(z)}{\gamma(z)})^{2}$. In Fig. \ref{Fig1} we plot
the space average mean exit time
$\tau_{_{ \textcolor{black}{SE}}}(z_{a})$  as a function of the
position of the virtual layer $z_{a}$. The continuous line is the result of scaling $\left(\frac{\tau_{_{ \textcolor{black}{SE}}}(z_{a})}{\tau^{bulk}_{_{ \textcolor{black}{SE}}}}\right)$ by the factor $\left[\frac{D(z_{a})}{D^{bulk}}\right] $ according to Eqs. (\ref{tau(a)}) and (\ref{eqavMFPT}).  The
quantity $[ D(z_{a})\,\tau_{_{ \textcolor{black}{SE}}}(z_{a})]$
was numerically evaluated  from Eq. (\ref{eqavMFPT}), using as input only
the equilibrium local particle density $\rho(z)$. As we can see, from Fig. \ref{Fig1} the scaling of the  \textcolor{black}{SE} and MD predicted by Eq. (\ref{eqScaling}) gives a very good numerical agreement.

\section{Conclusions}
Once the relaxation time of the persistence probability is identified as the space average of the mean first passage time, the  evaluation of the quantity $[ D(z_{a})\,\tau_{_{ \textcolor{black}{SE}}}(z_{a})]$ is easily obtained in terms of the equilibrium local particle density for a highly anisotropic dense fluid, next to an attractive surface. We have obtained a relationship, Eq. (\ref{eqScaling}), which shows the correspondence with the VLMD value, for a fixed width of the absorbing virtual layer. The simple scaling factor in terms of the anisotropic diffusion coefficient is physically meaningful, since $D(z)$ is introduced in the Langevin equation to account phenomenologically for the fluctuation--dissipation theorem.
\begin{acknowledgments}
This work was supported by Grant CDCHT-ULA-CVI-ADG-C09-95.
\end{acknowledgments}


\begin{thebibliography}{99}
 \bibitem{clor} Colmenares, P. J.; L\'opez, F.; Olivares--Rivas, W. {\it Phys. Rev. E} {\bf 2009}, 80, 061123.
 \bibitem{liu} Liu, P.; Harder, E.;  Berne, B. J. {\it J. Phys. Chem. B}, {\bf 2004}, 108, 6595.
 \bibitem{burska}Burschka, M. A.; Titulaer, U. M. {\it J. Stat. Phys.} {\bf 1981}, 25, 569.
 \bibitem{harris}Harris, S. {\it J. Chem. Phys.} {\bf 1981}, 75, 3103.
 \bibitem{razi}Razi Naqvi, K.; Mork K. J.; Waldenstr{\o}m, S. {\it Phys. Rev. Letts.} {\bf 1982}, 49, 304.
 \bibitem{thomas}Thomas, J. A.; McGaughey, J. H.  {\it J. Chem. Phys.}, {\bf 2007}, 126, 0347071.
\bibitem{gardiner}Gardiner, C. W. {\it Handbook of Stochastic Methods for Physics, Chemistry
  and the Natural Sciences}; Springer--Verlag: Berlin, 1985.
 \bibitem{ColmenaresPRE} Colmenares, P. J.; Olivares--Rivas, W.  {\it Phys. Rev. E} {\bf 1999}, 59, 841.
 \bibitem{SOVCLTheochem}Sulbar\'an, B.; Olivares--Rivas, W.; Villegas, J. C.; Colmenares, P. J.; L\'opez, F.   {\it J. Mol. Struct.: THEOCHEM}, {\bf 2006}, 769, 151.
 \bibitem{SulbaranCMP}Sulbar\'an, B.; Olivares Rivas, W.; Colmenares, P. J. {\it Cond. Matt. Phys.} {\bf 2005}, 8, 303.
\bibitem{israelachvili}Israelachvili J. {\it Intermolecular Surface Forces}, 2nd. Ed., Academic Press Ltd: London, 1992.
\bibitem{Travis&Gubbins}Travis, K. P.; Gubbins, K. E. {\it J. Chem. Phys.} {\bf 2000}, 112, 1984.
\bibitem{mcquarrie} McQuarrie D. A. {\it Statistical Mechanics} , Harper Row: New York, 1976.
\bibitem{risken} \textcolor{black}{H. Risken, The Fokker-Planck Equation. Methods of Solutions and Applications. Springer, 1984.}
\bibitem{ake} \AA kesson, T; J\"{o}nsson B; Halle B;  Chang D. Y. {\it Mol. Phys.} {\bf 1986},  57, 1105.
\bibitem{jonsson} \textcolor{black}{J\"{o}nsson B.; Wennestr\"{o}m H. in {\it Micelar Solutions and Microemulsions } edited by S. H. Chen and R. Rajagopalan; Springer--Verlag: New York, 1990.}
\bibitem{szabo}\textcolor{black}{Szabo A.; Schulten K.; Schulten Z. {\it J. Chem. Phys.} {\bf 1980}, 72(8), 4350.}

\end{thebibliography}
\end{document}